# Activity and Circadian Rhythm of Sepsis Patients in the Intensive Care Unit

Anis Davoudi, Duane B. Corbett, Tezcan Ozrazgat-Baslanti, Azra Bihorac, Scott C. Brakenridge, Todd M. Manini, Parisa Rashidi

*Abstract*— Early mobilization of critically ill patients in the Intensive Care Unit (ICU) can prevent adverse outcomes such as delirium and post-discharge physical impairment. To date, no studies have characterized activity of sepsis patients in the ICU using granular actigraphy data. This study characterizes the activity of sepsis patients in the ICU to aid in future mobility interventions. We have compared the actigraphy features of 24 patients in four groups: Chronic Critical Illness (CCI) sepsis patients in the ICU, Rapid Recovery (RR) sepsis patients in the ICU, non-sepsis ICU patients (control-ICU), and healthy subjects. We used several statistical and circadian rhythm features extracted from the patients' actigraphy data collected over a five-day period. Our results show that the four groups are significantly different in terms of activity features. In addition, we observed that the CCI and control-ICU patients show less regularity in their circadian rhythm compared to the RR patients. These results show the potential of using actigraphy data for guiding mobilization practices, classifying sepsis recovery subtype, as well as for tracking patients' recovery.

## I. INTRODUCTION

Sepsis is defined as an inflammatory body response to infection, with severe sepsis and septic shock being its more severe forms (1). Sepsis has a high prevalence rate of up to 30% in the Intensive Care Unit (ICU) (2). Sepsis prevalence has been increasing, possibly due to progressive aging of population, and the existence of more comorbidities (3-5). Sepsis can negatively affect health outcomes in ICU patients, including higher chance of mortality, longer length of stay in the ICU, higher chance of need for specialized care after discharge, and long-term decline in cognitive and functional abilities after discharge (3, 6-10). With higher rates of sepsis in the ICU, there is an increasing population of sepsis survivors that will be dealing with its consequences.

Sepsis patients, who do not die early, can be classified into two recovery subtypes: Chronic Critical Illness (CCI) and Rapid Recovery (RR). CCI type is defined as an ICU length of stay greater than or equal to 14 days with the evidence of persistent organ dysfunction. Patients with ICU length of stay of less than 14 days also qualify for CCI if they are discharged to another hospital, a long-term acute care facility, or to hospice, and demonstrate evidence of organ dysfunction at the time of discharge. Patients who do not meet the criteria for CCI or early death, are classified as RR (11, 12). CCI patients have a greater incidence of secondary infections (11).

Early detection of sepsis can significantly affect patient recovery rate (13-16). Current methods of diagnosis rely on physiological signals such as heart rate and core body temperature or laboratory tests such as procalcitonin (17). Despite these efforts, sepsis is still unrecognized and underreported (18). Besides diagnosing sepsis, identifying sepsis recovery subtypes can also allow for timely interventions that can reduce sepsis duration and severity.

While electronic health records data and several physiological signals have been used for sepsis detection and its risk prediction (19, 20), other information such as functional status and activity level have rarely been examined. Typically, questionnaire-based assessment tools are used for assessing patient activity and functional status in the ICU or after discharge (21, 22). However, these tools often introduce uncertainties such as recall bias and subjectivity. In recent years, actigraphy methods have been used in various studies for continuous, noninvasive, and objective assessment of activity over long periods (23, 24). Continuous and accurate activity measurement in the ICU can also guide mobilization interventions, and can lead to improved patient outcomes (25, 26).

In this study, we used actigraphy to characterize activity of sepsis patients in the ICU. We have examined activity patterns of CCI and RR sepsis patients in comparison with both non-sepsis ICU patients and healthy subjects. To our knowledge, this is the first study to use actigraphy data to characterize sepsis recovery subtypes. In addition, we compared the circadian rhythm of CCI and RR sepsis patients with both non-sepsis ICU patients and healthy subjects. Circadian rhythm is important for maintaining health in humans, and can be affected by both sepsis and length of ICU stay. To our knowledge, the effect of sepsis on circadian rhythm of physical activity has not been studied before, since it requires continuous activity measurement. Therefore, we examined patients' circadian rhythm to understand how sepsis affects the diurnal rhythmicity of physical activity for each sepsis recovery subtype, compared to non-sepsis ICU patients and healthy subjects.

The rest of the paper is as follows: we first explain the dataset description and analysis methods in section II, we will describe the results in section III, and finally we will discuss the results in section IV.

A. Davoudi and P. Rashidi are with the Biomedical Engineering Department, University of Florida, Gainesville, FL 32611 USA. (emails: anisdavoudi@ufl.edu, parisa.rashidi@bme.ufl.edu)

D. C. Corbett and T. M. Manini are with the Department of Aging and Geriatric Research, University of Florida, Gainesville, FL 32610, USA (emails: dcorbett@ufl.edu, tmanini@ufl.edu).

T. Ozrazgat-Baslanti and A. Bihorac are with the Department of Medicine, University of Florida, Gainesville, FL 32610 USA (emails: tezcan@phhp.ufl.edu, abihorac@ufl.edu).

S. C. Brakenridge is with the Department of Surgery, University of Florida, Gainesville, FL 32610 USA (email: scott.brakenridge@surgery.ufl.edu).

## II. METHODS

### A. Data Collection

We collected actigraphy data from 14 patients admitted to UF Shands hospital ICU between 04/2016-06/2017, and from 10 community-dwelling subjects. All subjects were consented to participate in the study prior to enrollment, and all procedures were approved by the University of Florida Institutional Review Board (IRB). Participants included sepsis ICU patients from both CCI and RR groups, non-sepsis ICU patients, and healthy subjects. The participants wore an ActiGraph GT3X on their dominant wrist. ActiGraph GT3X is an accelerometer unit used for continuous and noninvasive measurement of human physical movement. In this study, we used activity expressed in counts per minute. ICU patients wore the device for the duration of their stay in the ICU, and healthy subjects wore the device for two weeks. Five days of actigraphy data was used for analysis for all participants.

### B. Analysis

First, we removed days with non-weartime longer than one hour at a time. We used the vector magnitude activity counts calculated as in Equation (1) for features extraction. We calculated the average and confidence intervals for all patients in each group for visualizing the difference in activity patterns among different groups. We compared CCI and RR activity to both healthy subjects and control-ICU patients to examine the effect of ICU admission on patients' activity and circadian rhythm.

$$vector\ magnitude = \sqrt{x^2 + y^2 + z^2} \quad (1)$$

In Equation (1), x, y, and z are the activity counts in the three Cartesian basis vectors.

We extracted several statistical features as well as circadian rhythm features to summarize the 5-day activity data. We used these features to compare the distribution of features among the four participant groups. For features that did not have a normal distribution, we used nonparametric tests to describe the feature distributions, and to assess the differences among groups. All analysis were done using R (version 3.1.3) (27).

To extract the circadian rhythm features, we used a non-linear transformation of the traditional cosine curve, using the anti-logistic function in the sigmoidal family, as in Equation (2), (28). We extracted the following features on the activity's circadian rhythm from the fitted nonlinear model for each patient: *min, alpha, beta, phase, amplitude,* and *mesor*. *Min* is the minimum value of the fitted model. *Phase* is the time of day the fitted model's peak occurs. *Mesor* is the adjusted mean value. Amplitude is the difference between the minimum and maximum of the fitted model. *Alpha* determines whether the peaks of the model's fitted values are wider than their troughs, and *beta* determines whether the model rises and falls more steeply than a cosine curve. We performed the parameter estimation in two stages. In the first stage, the parameters of the traditional cosine curve were estimated by linear least squares projection of the data onto sine and cosine curves of 24h period. The linear model coefficients were then transformed in a non-linear manner into mesor (estimated by the mean in this case), amplitude and phase. Next, parameters of the extended cosine model were estimated using non-linear least squares, with the starting values of the parameters computed from mesor, amplitude, and phase of the best-fitting cosine curve (28).

$$l(c(t)) = \frac{\exp(\beta[c(t)-\alpha])}{1+exp(\beta[c(t)-\alpha])} \quad (2)$$

$$where, \quad c(t) = \cos\left(\frac{[t-\phi]2\pi}{24}\right).$$

## III. RESULTS

Table I shows the demographic characteristics of participants in each group. Inclusion criteria included age of 18 years or older. Participants' age and gender do not differ significantly between groups, except for RR group's age that was different from healthy subjects. Fig. 1 shows the average and confidence interval of the patients' activity in all four groups over five days. Healthy subjects' activity is significantly higher than the other three groups, as expected. Also, the RR group's activity is higher than both CCI and control-ICU groups. However, CCI group and control-ICU group have similar activity over five days. Another difference between the groups is that healthy subjects have a clear circadian rhythm, while RR subjects have a less distinct circadian rhythm. The CCI and control-ICU subjects do not show any rhythmicity and difference between the daytime and nighttime activity.

Table II shows the distribution of statistical features for the four groups. All features were statistically different among the four groups (except for the start of 10-hour maximum activity window). The extracted features are different between healthy subjects and all the other groups (Fig.1). Half of the features are statistically different between the RR group and all other groups, with a significant gap between RR group and other groups, as seen in Fig.1. None of the features were statistically different between the CCI and non-sepsis ICU patients (Fig.1).

We also visualized the differences between the circadian rhythm of the four groups' activity data. Fig. 2 shows the fitted non-linear model for one example patient per each group (other patients are not depicted due to space constraints). As in Fig. 1, the healthy subject and RR patient show a clearer rhythmicity in their activity, compared with the control-ICU and CCI patients. Table III shows the distribution of circadian features for all patients, extracted using a sigmoidally transformed cosine function. For this dataset, only the amplitude feature was significantly different between the four groups (p<0.001).

## IV. DISCUSSION

We showed that there are significant differences in the activity profiles of CCI and RR sepsis patients during their stay in the ICU. As expected, actigraphy features of CCI, RR, and non-sepsis ICU patients were significantly different from healthy subjects. This concurs with previous non-actigraphy studies of functional status in sepsis and ICU patients. Among the circadian rhythm features, amplitude was the only feature that was significantly different among the four groups.

However, a larger population might show a larger difference in the distribution of other circadian rhythm features as well. Differences in physical activity of the patients may become more evident over longer stay in the ICU, and also due to the muscle deterioration caused by sepsis over time. This is the first study that uses actigraphy methods to objectively and continuously measure sepsis patients' activity during their ICU stay, and to compare it to non-sepsis ICU patients and healthy subjects. The significant differences in actigraphy features show that they can potentially be used for automatic detection of sepsis severity and recovery subtype.

One main limitation of the study was the small number of participants in each group, which might have contributed to inability to capture potential differences in features among the four groups. The small sample size also resulted in a limited age range, with the sample generally consisting of older adult patients. Another limiting factor was that the participants were not matched according to their comorbidity and primary diagnoses. This may contribute to intragroup diversity when unaccounted for.

In the future, we will investigate the discriminating power of more diverse actigraphy features within a larger and more diverse population.

TABLE I Characteristics of study cohort. Control-ICU: non-sepsis ICU patients, CCI: chronic critical illness, RR: rapid recovery, control-healthy: healthy subjects. N: number of patients. Un: Unavailable.

| Variable | Control-ICU, | CCI, N=5 | RR, N=6 | Control-healthy, | P[a] |
|---|---|---|---|---|---|
| Age, median (25%, 75%) | 68 (59,79) | 63 (51, | 58 (52.3, | 67 (66,67.8) | 0.152 |
| Gender, number of female (%) | 2 (0.66) | 2 (0.4) | 4 (0.66) | 4 (0.4) | 0.568 |
| Race, number of white (%) | 3 (100) | 5 (100) | 6 (100) | Un | 1 |
| Primary Diagnosis Group, number (%) | | | | | 0.730 |
| - Certain infectious and parasitic diseases | 0 (0) | 2 (40) | 2 (33.3) | Un | |
| - Diseases of the digestive system | 3 (100) | 2 (40) | 2 (33.3) | Un | |
| - Malignant neoplasms of female genital organs | 0 (0) | 0 (0) | 1 (16.7) | Un | |
| - Injury, poisoning and certain other consequences of external causes | 0 (0) | 0 (0) | 1 (16.7) | Un | |
| - Diseases of the nervous system | 0 (0) | 1 (20) | 0 (0) | Un | |

a. Kruskal-Wallis analysis of variance by ranks.

TABLE II Distribution of statistical features extracted from activity data of all groups. Control-ICU: non-sepsis ICU patients, CCI: chronic critical illness, RR: rapid recovery, control-healthy: healthy subjects. Data are median and interquantile range (25%, 75%) values. [a]: Kruskal-Wallis analysis of variance by ranks, [b]: significantly different from control-healthy patients (p<0.01), [c]: significantly different from CCI patients (p<0.05), [d]: significantly different from RR patients (p<0.05), [e]: significantly different from control-ICU patients (p<0.05). N: number of patients. M10: activity over the 10-hour window with maximum activity. L5: activity over the 5-hour window with minimum activity. RA: relative amplitude (defined as $\frac{M10-L5}{M10+L5}$). RMSSD: root mean square of sequential of sequential differences. SD; standard deviation.

| Variable | Control-ICU, N=3 | CCI, N=5 | RR, N=6 | Control-healthy, N=10 | p[a] |
|---|---|---|---|---|---|
| Mean of activity of the whole day | 31.1 (22.8, 128)[b,d] | 140 (83.6, 168)[b,d] | 428 (313, 551)[b,c,e] | 1328 (1062, 1742)[d,e,f] | <0.001 |
| Standard deviation of activity of the whole day | 137 (113, 308)[b,d] | 346 (249, 386)[b,d] | 701 (586, 990)[b,c,e] | 2049 (1663, 2107)[d,e,f] | <0.001 |
| M10 | 25513 (18571, 113443)[b,d] | 126226 (67371, 151684)[b,d] | 317549 (252610, 492412)[b,c,e] | 1398892 (1083188, 1685662)[d,e,f] | <0.001 |
| Time of M10 | 437 (369, 572) | 491 (405, 568) | 602 (517, 616) | 513 (418, 580) | 0.585 |
| L5 | 984 (709, 11603)[d] | 12014 (10729, 17500)[b,d] | 59283 (43613, 73749)[c,e] | 23062 (21363, 26593)[d] | 0.002 |
| Time of L5 | 389 (257, 515) | 696 (571, 792)[b] | 298 (166 ,436)[b] | 49.9 (26.7, 78.2)[d,e,f] | 0.002 |
| RA | 0.84 (0.82, 0.90) | 0.78 (0.78, 0.86)[b] | 0.74 (0.68, 0.81)[b] | 0.96 (0.95, 0.97)[d,e] | 0.001 |
| RMSSD | 143 (123, 327)[b,d] | 364 (253, 369)[b,d] | 760 (593, 912)[b,c,e] | 1297 (1149, 1447)[d,e,f] | <0.001 |
| RMSSD/SD | 1.11 (1.09, 1.13)[b] | 1.05 (1.05, 1.07)[b] | 0.97 (0.91, 1.06)[b] | 0.66 (0.62, 0.69)[d,e,f] | <0.001 |
| Number of immobile minutes | 1239 (1040, 1287)[b,d] | 1046 (1029, 1102)[b,d] | 651 (621, 682)[b,c,e] | 564 (517, 590)[d,e,f] | <0.001 |

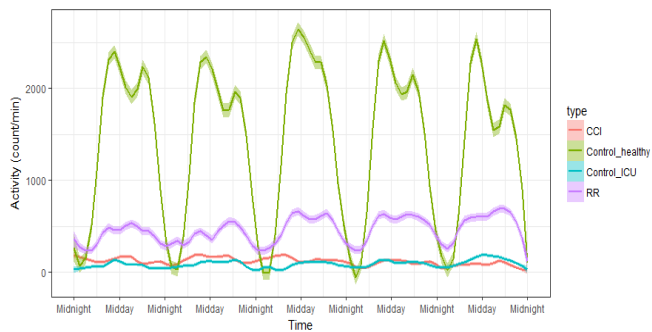

Figure 1 Average activity of four groups of patients –CCI: chronic critical illness patients, RR: Rapid Recovery patients, control-ICU: non-sepsis ICU patients, control-healthy: healthy subjects. For all groups, confidence interval is shown by a transparent band around the average values.

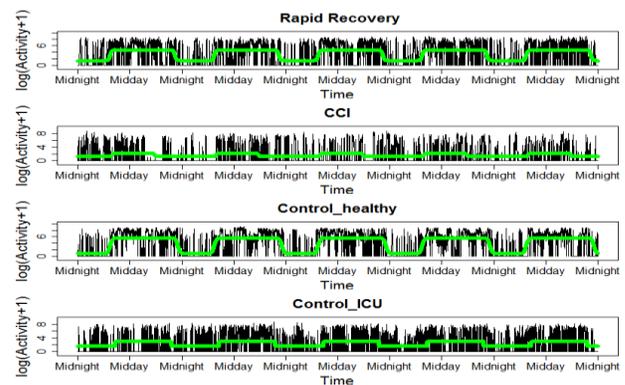

Figure 2 Nonlinear fitted model (green) used for circadian rhythm features extraction, for one patient per group. CCI: chronic critical illness patients, control-ICU: non-sepsis ICU patients, control-healthy: healthy subjects.

Table III Distribution of statistical features extracted for activity data for all groups. Control-ICU: non-sepsis ICU patients, CCI: chronic critical illness, RR: rapid recovery, control-healthy: healthy subjects. Data are median and interquartile range (25%, 75%) values. [a]: Kruskal-Wallis analysis of variance by ranks, [b]: significantly different from control-healthy patients (p<0.01), [c]: significantly different from CCI patients (p<0.05), [d]: significantly different from RR patients (p<0.05), [e]: significantly different from control-ICU patients (p<0.05). N: number of patients.

| Variable | Control-ICU, N=3 | CCI, N=5 | RR, N=6 | Control-healthy, N=10 | p[a] |
|---|---|---|---|---|---|
| Min | 0.55 (0.39, 1.02) | 0.99 (0.67, 1.06) | 1.94 (0.84, 2.61) | 0.79 (0.49, 1.13) | 0.475 |
| Amplitude | 0.72 (0.44, 1.07)[c] | 0.77 (0.60, 1.08)[c] | 0.81 (0.61, 2.43)[c] | 5.64 (4.97, 6.04)[d,f] | <0.001 |
| Phase | 6.21 (4.71, 10.66) | 13.09 (12.81, 15.71) | 15.41 (13.67, 16.14) | 14.75 (13.78, 15.23) | 0.281 |
| Alpha | -0.13 (-0.15, 0.43)[c] | 0.20 (-0.65, 0.27) | -0.50 (-0.74, -0.04) | -0.44 (-0.47, -0.35)[f] | 0.249 |
| Beta | 27.34 (16.02, 291.46) | 51.77 (40.81, 56.69) | 48.13 (18.91, 78.86) | 14.47 (8.16, 33.93) | 0.398 |